**Effect of clouds on emission spectra for Super Venus**

Paulina Wolkenberg[1] and Diego Turrini[1,2]

[1]Istituto di Astrofisica e Planetologia Spaziali (IAPS) – Istituto Nazionale di Astrofisica (INAF), Via del Fosso del Cavaliere 100, Rome, 00133, Italy

[2]Osservatorio Astrofisico di Torino (OATo) – Istituto Nazionale di Astrofisica (INAF), Via Osservatorio 20, Pino Torinese, 10025, Italy

**Corresponding author**:

Paulina Wolkenberg

IAPS – INAF

Via del Fosso del Cavaliere 100

Rome, 00133, Italy

e-mail: paulina.wolkenberg@inaf.it

ORCID 0000-0001-6769-3719

**Abstract**

We report a model study on the effects of clouds on emission spectra of super-Venus planets. Our goal is to assess possible ways to identify characteristic spectral features due to clouds. We show that it is possible to distinguish an impact of $H_2SO_4$ clouds on the $CO_2$ absorption band at 4.8 μm for temperature profiles with and without a thermal inversion. The thermal inversion can help to distinguish the signal from high altitude clouds (85 km, ~1 mbar). Featureless emission spectra are found for high altitude clouds (85 km, ~1 mbar) with temperature profile without thermal inversion. More spectral features appear in the emission spectra with decreasing cloud top altitudes. The compactness of clouds has an inverse effect on emission spectra than cloud top altitudes. Small cloud scale heights reduce the signal and the $CO_2$ absorption bands become flat.

**1. Introduction**

Clouds in the Solar system are ubiquitous. They are found in most of the planets and the moons with atmospheres, like Titan. Thus, it is reasonable to anticipate that they can also form in atmospheres of exoplanets. Our purpose is to characterize the possible impact of clouds on emission spectra in eclipse for smaller planets, with sizes typical of Super-Earths. We aim to find characteristic features of clouds for two types of atmospheric temperatures, one of them including inversion, and their effect on emission spectra. We also discuss the possibility of detection by future facilities.

Atmospheric models with clouds were proposed to explain and to be consistent with observations of hot Jupiter planets such as HD 209458b and XO-1b (Knutson et al., 2009, Sing et al., 2016, Tsiaras et al., 2018, Pinhas et al., 2019, Welbanks et al., 2019). However, cloud characterization is a difficult task so far because of sparse and low resolution data obtained from different instruments that have a variable sensitivity of measurements.

The difficulty of cloud detection and its characterization, particularly for smaller planets, is caused by the associated featureless spectrum in transit, which can provide information either about the absence of a gaseous atmosphere or the presence of clouds at high altitude (Benneke and Seager, 2013, Kreidberg, 2014, De Wit et al., 2016, de Wit et al., 2018). Clouds and hazes increase the reflected flux in the visible and near-infrared region of the eclipse spectrum (Sudarsky et al., 2003; Wakeford & Sing, 2015) and significant absorption features can be present (Ariel Definition Study Report, Fig.2-9). On the other hand, many transmission spectra of hot Jupiters are dominated by the strong optical Rayleigh/Mie scattering from high-altitude aerosol particles. The strong optical slope in the visible range of the transmission spectrum is associated with more prominent condensate vibrational mode features in the infrared range due to clouds being composed of small sub-micron sized particles. However, clouds and hazes can effectively shadow any absorption features of different components coming from the deeper atmosphere in transmission spectra in the optical and near infrared range, for example from water vapor (Wakeford & Sing, 2015).

The weak signal of water vapor in the spectra of hot-Jupiters was initially considered as a result of low abundance of this atmospheric component. In turn, this would mean that the abundance of water in the protoplanetary disk at the planet's formation location decreased. However, the comparative study for ten exoplanets suggested that the reason of low signature by water vapor can be due to the occurrence of high altitude clouds or haze. This study also rejected the hypothesis about the primordial water depletion during formation (Sing et al., 2016). However, Welbanks et al. (2019) claim that the low abundances of $H_2O$ found in atmospheres of hot Jupiters indicate different formation pathways with respect to the solar system giants.

Vasquez et al. (2013) found that clouds composed of water ice (high level) or liquid water (low level) have a strong effect on the intensity and shape of bands of $CO_2$, $N_2O$, $H_2O$, $CH_4$ and $O_3$ in the high-resolution flux spectra, leading to the reduction of their detection mostly for planets orbiting the F-type star. On the other side, they showed that clouds can be also useful to detect $CH_4$ and $O_3$, which are difficult to distinguish in the cloud-free atmosphere.

Contrary to cloudy models, cloud-free, hydrogen-dominated atmospheres generate absorption features in the transmission spectrum that are several times larger than those expected for atmospheres containing large mass fractions of water vapor, carbon dioxide, carbon monoxide, methane or nitrogen due to the lower mean molecular mass and resulting larger scale height (Miller-Ricci et al., 2009; Miller-Ricci and Fortney, 2010). However, the interpretation of measurements of several exoplanets is still ambiguous because observations agree with two types of atmospheric models, namely, one composed of hydrogen with high-altitude clouds and a second with an atmosphere with high molecular weight such as water vapor in case of the super-Earth GJ 1214b (Kreidberg et al., 2014). Other examples of small planets for which the interpretation of spectral observations is still ambiguous are: Trappists (De Wit et al., 2016, 2018), K2-18b (Tsiaras et al., 2019, Benneke et al., 2019, Madhusudhan et al., 2020), LHS-1140b (Edwards et al., 2021), GJ-1132b (Swain et al., 2021, Mugnai et al., 2021).

High-altitude clouds are suggested to explain a flat transmission spectrum of the super-Earth GJ 1214b taken by the wide field camera 3 on the Hubble Space Telescope (Berta et al., 2013). However, this hypothesis requires the optically thick cloud layer to extend up to 10 mbar or particles with the diameter size approaching 1 μm. Thus they conclude that the flatness of spectrum is rather due to an atmosphere with a high mean molecular weight. The transmission spectra for the super-Earth GJ 1214b calculated using photochemical models point out the presence of clouds or hazes and the absence of methane in its atmosphere. This suggests that the atmosphere of GJ 1214b is hydrogen-rich with efficient photolysis of methane and carbon-rich hazes (Miller-Ricci Kempton et al., 2012). The transit spectra of the super-Earth GJ 1214b exclude the cloud-free atmosphere composed of hydrogen. Too high uncertainty of measurements does not allow to explain whether the featureless spectrum is due to high-altitude clouds or the presence of a high mean molecular mass atmosphere (Benneke and Seager, 2013).

After analysis of observations of the transit of GJ 1214b performed by INT (Isaac Newton Telescope), MPI/ESO telescope, NOT (Nordic Optical Telescope (NOT) and WHT (William Herschel Telescope), de Mooij et al. (2012) found a tentative increase in the planet-to-star size ratio at the shortest wavelength, which could be interpreted as an increase of planet-size due to Rayleigh scattering. This implied that the atmosphere of GJ 1214b was probably composed of low mean molecular weight gas such as hydrogen/helium with large scale height. The best agreement with data was observed for an atmosphere with a sub-solar metallicity and low content of methane with a grey cloud layer at a pressure of 0.5 bar. However, they concluded that it required more observations and better estimations of spectral stellar spot behavior at optical wavelengths to confirm this statement.

Moreover, after more precise analysis of three transits of GJ 1214b in the blue visible channel, de Mooij et al. (2013) excluded the hypothesis on the increase of planet-to-star radius ratio due to Rayleigh scattering, which could be the indication of the low molecular weight atmosphere with the large scale height. The ratio derived from three transits at the blue optical range was found very similar. Both models of an atmosphere with hydrogen and clouds, and an atmosphere with water vapor fit the data, although the best agreement was found for an atmosphere with small scale height, high molecular weight such as water vapor. Morley et al. (2013) also suggested that reasons of a relatively featureless spectrum of super-Earth GJ 1214b could be the water-rich atmospheres or the hydrogen/helium composition with clouds.

Morley et al. (2013) presented a model with two types of clouds that might explain the flat transmission spectrum. They used a model with KCl and ZnS cloud composition in chemical equilibrium and, separately, the hydrocarbon haze layer as a result of photochemistry. Both models matched the observations provided that the equilibrium cloud sedimentation efficiency was low and the particle sizes of the haze layer were from 0.01 to 0.25 micron. Discrimination between cloudy and water rich atmospheres was provided by Kreidberg et al. (2014) for super-Earth GJ 1214b and Knutson et al. (2014) for the Neptune-mass exoplanet GJ 436b. However, while Kreidberg et al. (2014) claimed that observations confirmed the existence of the atmosphere with clouds, they considered only observations from 1.1 – 1.7, micron that do not take into account the visible part of the spectrum.

## 1.1. Thermal inversion and clouds

Knutson et al. (2009) provide a classification of atmospheres of hot Jupiter exoplanets. They found that the emission spectra of some hot Jupiter planets are consistent with the one-dimensional cloud-free atmosphere model (Knutson et al., 2009, and references therein). The infrared emission spectra of these planets are characterized by strong absorption bands of CO and $H_2O$. However, there are other hot Jupiter planets such as HD 209458b and XO-1b that are not in agreement with the cloudless atmosphere. Knutson et al. (2008) found a temperature inversion at high altitude in the atmosphere of HD 209458b comparing the planet's emission at shorter wavelengths for standard cloudless and cloudy atmosphere models. In order to be consistent with observed spectra of HD 209458b and XO-1b in the produced model emission spectra, the cloudy model atmosphere with a temperature inversion between 0.1 and 0.01 bars and water bands in emission instead of absorption were required (Knutson et al., 2009). Madhusudhan and Seager (2009) successfully applied their retrieval algorithm for temperature and abundances to two transiting exoplanets, HD 189733b and HD 209458b. They constrained the molecular abundances of $H_2O$, $CO$, $CH_4$ and $CO_2$ in the atmosphere of HD 189733b and confirmed an absence of the thermal inversion. They confirmed the dayside thermal inversion in the atmosphere of HD 209458b. However, the thermal inversions are still debated for HD-209458 b (Schwarz et al., 2015, Line et al., 2016).

The water emission features at 4.5 and 5.8 micron could indicate the temperature inversion in the upper atmosphere. Earlier spectrum by IRS (Spitzer Infrared Spectrograph) in the 7.5 – 13.2 μm range showed no evidence of a water absorption at shortest wavelengths. To the contrary, the spectrum contained two emission features, one of them identified as an emission from silicate clouds (Richardson et al., 2007). The presence of high altitude clouds in the atmosphere of HD 209458b could explain the sodium absorption weaker than predicted in the planet's transmission spectrum (Charbonneau et al., 2002) but there was currently no definitive evidence for the presence of such clouds.

The Spitzer Space Telescope infrared observations have found that several giant planets have a thermal inversion at high altitude (Burrows et al., 2007; Knutson et al., 2008; Fortney et al., 2006; Burrows et al., 2008; Knutson et al., 2009; Machalek et al., 2008). Observations of the TrES-4 planet suggested that highly irradiated planets are more likely to have thermal inversions (Knutson et al., 2009). Thermal inversions were also detected in the measurements by HST (Madhusudhan et al., 2010, Haynes et al., 2015, Sheppard et al., 2017, Kreidberg et al., 2018, Evans et al., 2018, Evans et al., 2020, Edwards et al., 2020, Pluriel et al., 2020, Changeat et al., 2021). However, the reason of thermal inversion origin is still unknown. Among competing theories, one with an optical absorber is mentioned.

Spiegel et al. (2009) considered that TiO and VO could be a potential reason of the thermal inversion in the upper atmosphere of giant exoplanets. Molecules such as TiO and VO are strong optical absorbers. If such chemical species are present at high altitude, they lead to a thermal inversion in which the temperature increases with altitude. This is also the case of ozone in the Earth stratosphere. However, after model calculations on 5 giant exoplanets including HD 209458b, HD 149026b, TrES-4, OGLE-TR-56b and WASP-12-b they excluded VO as a producer of the thermal inversion. However, the recent studies on HD 209458b showed no thermal inversion in the atmosphere (Schwarz et al., 2015, Line et al., 2016). The TiO hypothesis is maintained provided that macroscopic mixing such as turbulent diffusion occurs. However, the TiO gas existence in the upper atmosphere requires large values of eddy diffusion coefficient, thus the hypothesis becomes very problematic as concluded by Spiegel et al. (2009). Without the mixing process the TiO could condense and settle more strongly than the primary constituent like molecular hydrogen and concentrate at deep layers leading to the depletion of TiO at high altitudes, in turn making it ineffective in generating the thermal inversion.

A stratospheric thermal inversion above 10 mbar was also found in the atmosphere of GJ 1214b due to the heating by ZnS clouds used in a 3D General Circulation Model including cloud radiative effects (Charnay et al., 2015). However, the model has shown that clouds can increase the reflected radiation by increasing the planetary albedo, leading to the atmospheric cooling below 1 mbar.

### 1.2. Cloud chemical compositions

Clouds can be composed of different species dependently on planets. Hot Jupiters allow for the existence of heavy molecules in the upper atmosphere because of their high temperatures. Spiegel et al. (2009) suggested TiO and VO as components of clouds. The enstatite $MgSiO_3$ is assumed to be condensed at the stratospheric temperatures of HD 189733b (Barstow et al., 2014) and in Kepler-7b (Demory et al., 2013). Another component of clouds such as MnS is suggested to occur in the atmospheres of hot Jupiters (Barstow et al., 2014; Morley et al., 2013). Different compositions of clouds including KCl/ZnS and the formation of a hydrocarbon haze layer were taken into account in modeling GJ 1214b atmosphere (Morley et al., 2013; Kreidberg et al., 2014; Charnay et al., 2015) to study its relatively featureless transmission spectrum.

The atmosphere of warm sub-Neptune (super-Earth) GJ 1214b with the presence of high and thick clouds or haze was simulated to reproduce the featureless transit spectrum (Charnay et al., 2015). The atmosphere was dominated by hydrogen with clouds composed of KCl and ZnS molecules. However, $Na_2S$ and other iron and silicates clouds could occur in the deeper atmosphere and be removed (Morley et al., 2013). Many chemical molecules such as silicates ($Mg_2SiO_4$, $MgSiO_3$, $SiO_2$, $Fe_2SiO_4$, $FeSiO_3$), aluminum oxides ($Al_2O_3$, $MgAl_2O_4$), iron oxides ($FeO$, $Fe_2O_3$), titan oxides ($CaTiO_3$, $TiO_2$), sulphides ($Na_2S$, $MnS$, $ZnS$), chlorides ($NaCl$, $KCl$) were tested as potential components of clouds in hot Jupiters with different size distributions (Wakeford and Sing, 2015).

Ground-based transmission spectra of the hot Jupiter HD 189733b from 0.94 – 2.42 micron revealed the presence of $H_2O$ hazes (Danielski et al., 2014). Analysis of reflection spectra of HD 189733b further revealed the presence of enstatite ($MgSiO_3$) clouds in the atmosphere (Barstow et al., 2014). Small particles less than 0.1 micron and 50 ppmv of Na were responsible for the best fit to the STIS data. However, the analysis of reflection spectra was not able to exclude larger size particles due to degeneracy. Furthermore, the authors of the latter analysis suggested to use cloud-free model atmospheres to retrieve a temperature profile because of poor sensitivity of thermal emission spectrum on the inclusion of clouds.

Ehrenreich et al. (2014), in studying GJ 3470b from observations taken by WFC3 camera on Hubble Telescope, used different atmospheric models to find the best fit with data. Among the considered models they found that only a pure hydrogen atmosphere with clouds and water vapor as a trace gas with less than 1 ppm matched the observations. Other models gave some inconsistency in the spectrum for wavelengths less than 1 micron and in the infrared region.

Clouds can be detected from transmission and emission spectra and their spatial and temporal distributions can be derived from the reflected light phase curves in the visible range (Webber et al., 2015). Clouds and their distribution across the planetary surface can have a large impact on observed signal (Karalidi et al., 2012), introducing complexity and uncertainty in the interpretation of planetary light curves (Fuji et al., 2011) as well as daily variations in the planetary signal. As a result, the interpretation of the observations, both in terms of atmospheric thermal profile and spectra (Marley et al., 2012), can differ with respect to cloud coverage and variability. The planetary albedo model used for Kepler-7b observations of phase curves showed that half of its dayside is covered by clouds with small sizes at high altitudes in the atmosphere (Webber et al., 2015). The two-column radiative-convective model of the atmospheres of Earth-like tidally locked planets illustrated the day-night thermal emission contrast that in turn can help to distinguish cloud models (Yang&Abbot, 2014).

Besides of temporal and spatial distribution of clouds, De Kok et al. (2011) pointed out that knowledge of the scattering properties of the clouds is essential during a simulation of thermal emission spectra from direct infrared observations or from secondary eclipse measurements. Scattering clouds cause the variation of gas absorption features dependently on particle size. This can lead to the wrong estimation of retrieved atmospheric properties. The radiative transfer model used by de Kok et al. (2011) showed that clouds with small particles influence thermal emission spectra significantly at short wavelengths and cold temperatures. The simulations of transit, emission and reflection spectra for the cloudy hydrogen-dominated atmosphere of GJ 1214b was consistent with Hubble Space Telescope (HST) observations provided that cloud particle radii were around 0.5 μm, small enough to loft to the upper atmosphere (Charnay et al., 2015).

According to data provided by HST, observations of wing steepness and relative depths of absorption features in moderate-resolution near-infrared transmission spectra (R~100) can unambiguously discriminate cloudy mini-Neptunes and volatile-dominated worlds (Benneke and Seager, 2013). Lowering of uncertainties by a factor of ~3 in the spectral transit depth measurements would allow to distinguish a cloudy hydrogen-rich and cloud-free water vapor atmosphere of the super-Earth GJ 1214b. More precise moderate-resolution transmission spectroscopy can detect near-infrared absorption features and determine directly the mean molecular weight. These observations should allow to unambiguously distinguish also between cloudy and high-atmospheric-metallicity in GJ 436b's atmosphere (Knutson et al., 2014).

The Rayleigh scattering by small particles or molecules with sizes equivalent to the radiation wavelength increases with decreasing wavelength. Thus high signal in the blue channel could be an evidence for the existence of small cloud particles in the atmosphere. The clouds with submicron sized grain particles are suspected to be responsible for the Rayleigh slope in the transmission spectra in the atmosphere of HD 189733b (Wakeford and Sing, 2015) because the vibrational mode absorption features in the infrared region are only observed for this distribution. Particles larger than 1 μm produce completely flat and featureless transmission spectra in the optical and infrared spectral range (Wakeford and Sing, 2015).

1.3. **Cloud characterization with future observational facilities**

In future observatories, such as the James Webb Space Telescope (JWST) and Ariel, by considering transmission along with emission spectra, characterization of exoplanets' atmospheres will be available with a better accuracy. JWST is a telescope with the 6.5 m of diameter to be launched in 2022. It will measure a signal from a very large spectral range from 0.6 to 28 μm. Three instruments such as Near-Infrared Imager and Slitless Spectrograph (NIRISS), Near-Infrared Camera (NIRCAM) and Near-Infrared Spectrograph (NIRSPEC) will observe the spectral range from 0.6 to 5 μm with different spectral resolutions. Mid-Infrared instrument (MIRI) composed of a camera and a spectrograph with different spectra resolutions respectively will observe objects from 5 to 28 μm. It will be the first instrument in space which will be able to measure the signal from exoplanets of various masses and temperature down to temperate rocky planets in the mid-infrared range.

However, it is expected that only 20% of its observation time will be dedicated to transiting exoplanet programs (Ariel Definition Study Report, 2020).

Ariel, the M4 mission of the Cosmic Vision program of the European Space Agency deemed to launch in 2029, will systematically observe the atmospheres of hundreds of exoplanets ranging from super-Jovian gas giants to super-Earths (Tinetti et al. 2018, Zingales et al. 2018, Turrini et al. 2018, Edwards et al. 2019). The spectral ranges from visible to 7 μm for Ariel (Tinetti et al. 2018) and to 28.3 μm for JWST (Wright et al., 2008; Greene et al., 2016) and the better resolution compared to HST and Spitzer telescopes can be helpful to break down the degeneracy for molecule abundances, temperature profiles and characterization of clouds. Ariel will be able to recognize a cloudy atmosphere with a low mass weight component and atmosphere with a dominant high molecular weight by means of repeated time series of eclipse and transit measurements with the high resolving power in its Tier 2 Deep Survey (Tinetti et al. 2018). Thanks to the simultaneous observations in the whole considered spectral range by Ariel (Tinetti et al. 2018, Edwards et al. 2019), particularly in the visible and near infrared region of spectrum, we will be able to identify the cloud occurrence and to investigate its chemical composition, as well as spatial and temporal distributions.

Clouds on super Earths, however, are expected to be very hard to detect and characterize even by such incoming future observatories. Furthermore, in many works so far (Greene et al., 2016, Edwards et al., 2020) clouds are not assumed during simulations of emission spectra in eclipse for different exoplanets. Greene et al. (2016) claim that clouds may have a lessened effect on emission than in transmission spectra due to their reduced effective optical depth. However, in their conclusion, it is still unclear if clouds suppress the spectral features of molecules occurring in the atmosphere of exoplanets. Here, we will try to answer to this issue by using the example with the atmosphere composed of $CO_2$ with $H_2SO_4$ clouds, modeled on the template of Venus' atmosphere.

Specifically, our main objective in this work is to study the characteristic spectral features due to clouds in the atmospheres of super-Earths planets with Venus-like atmospheres (super-Venus planets in the following). Assuming that the main component of the atmosphere is $CO_2$, we investigate the effects of $H_2SO_4$ clouds on emission spectra in eclipse. We study the impacts of different cloud top altitudes and scale heights on the intensities of signal and on detections of $CO_2$ absorption bands at 4.3 and 4.8 μm. We assume two different temperature profiles, one of them including a thermal inversion. As a result of this study we find that $CO_2$ spectral features allow to distinguish the impact of clouds on the $CO_2$ absorption band at 4.8 μm for the two temperature profiles we considered.

As we will show, a sensitivity of 1 ppm is likely required for the characterization of cloudy super-Venus planets. While this level of precision may be difficult to achieve with space facilities like

JWST and Ariel, future facilities capable of systematically reaching the required level of sensitivity will allow to explore in more details this challenging class of exoplanets.

## 2. Methods

In our study we focus on the investigation of smaller planets (radii up to 4.25 of Rearth) with Venus-like atmospheres dominated by high-Z elements. For the purpose of this exercise, such planets could be either Venus-like super-Earths or small giant planets that lost most of their atmospheric H/He. In this exploratory study our aim is to assess whether we can take advantage of the higher sensitivity and spectral coverage of future observatories to draw additional constraints on the characteristics of clouds in the atmospheres of such planets. We focus on simulation of emission spectra in the spectral range of the two $CO_2$ absorption bands at 4.3 and 4.8 μm with clouds at different altitudes.

The emission spectra are calculated according to the formula (Kreidberg L., 2018):

$$\frac{F_p}{F_s} = \frac{I(T,\nu)}{B(T_s,\nu)} \cdot \frac{R_p^2}{R_s^2} \quad (1)$$

$F_p/F_s$ – a ratio of fluxes from planet and star

$I(T,\nu)$ – the spectral radiances at given vertical temperature profile

$B(T_s,\nu)$ – the Planck function at $T_s$ (temperature of star)

$R_p$ – planetary radius

$R_s$ - radius of star

The contribution of reflected light is omitted because we expect it not being important, considering that for Venus the spectral spherical albedo is around 0.05 in the range from 3.9 to 5.1 micron (Titov, et al., 2007).

Host stars which are similar to Trappist 1 and have sizes of the order of the radius of Jupiter ($R_J$) or 0.1 Solar radii are the best candidates to have a high signal in emission spectra. We therefore consider a planetary system around an M8 star having a planet with radius around 4.25*$R_E$. The temperature of our template star is assumed as 2560 K like Trappist 1. The planet is assumed to be similar in atmospheric composition to Venus. The atmosphere of this template Super-Venus will be made mainly of $CO_2$ and $H_2SO_4$ clouds extending from 50 to 85 km of altitude. The clouds are characterized by the top altitude $zt$ and scale height $H$ that describes their compactness with altitude. A detailed description of vertical and size distribution for clouds can be found in Wolkenberg et al. (2018). In order to calculate the emission spectra, we use a radiative transfer model described in Wolkenberg et al. (2018) and Garcia-Munoz et al. (2013). Here, we summarize main points of this model.

Our model of atmosphere is mainly composed of $CO_2$ (96.5%) and $N_2$ (3.5%). The radiative transfer model accounts for the multiple scattering by clouds (Stamnes, 1988). The gas line parameters are taken from HITRAN 2008 database (Rothman et al., 2009). The Voigt line shape is used with sub-Lorentzian correction for line wings according to Winters (1964). We assume a cut off with 200 $cm^{-1}$ and the sub-Lorentzian modification beyond 5 $cm^{-1}$. Our cloud model is a log normal particle size distribution with mode 2 corresponding with $r_{eff}$ = 1.09 μm with variances $v_{eff}$ = 0.037. The cloud chemical composition with the 84.5% of $H_2SO_4$ by weight in the solution with $H_2O$ is assumed.

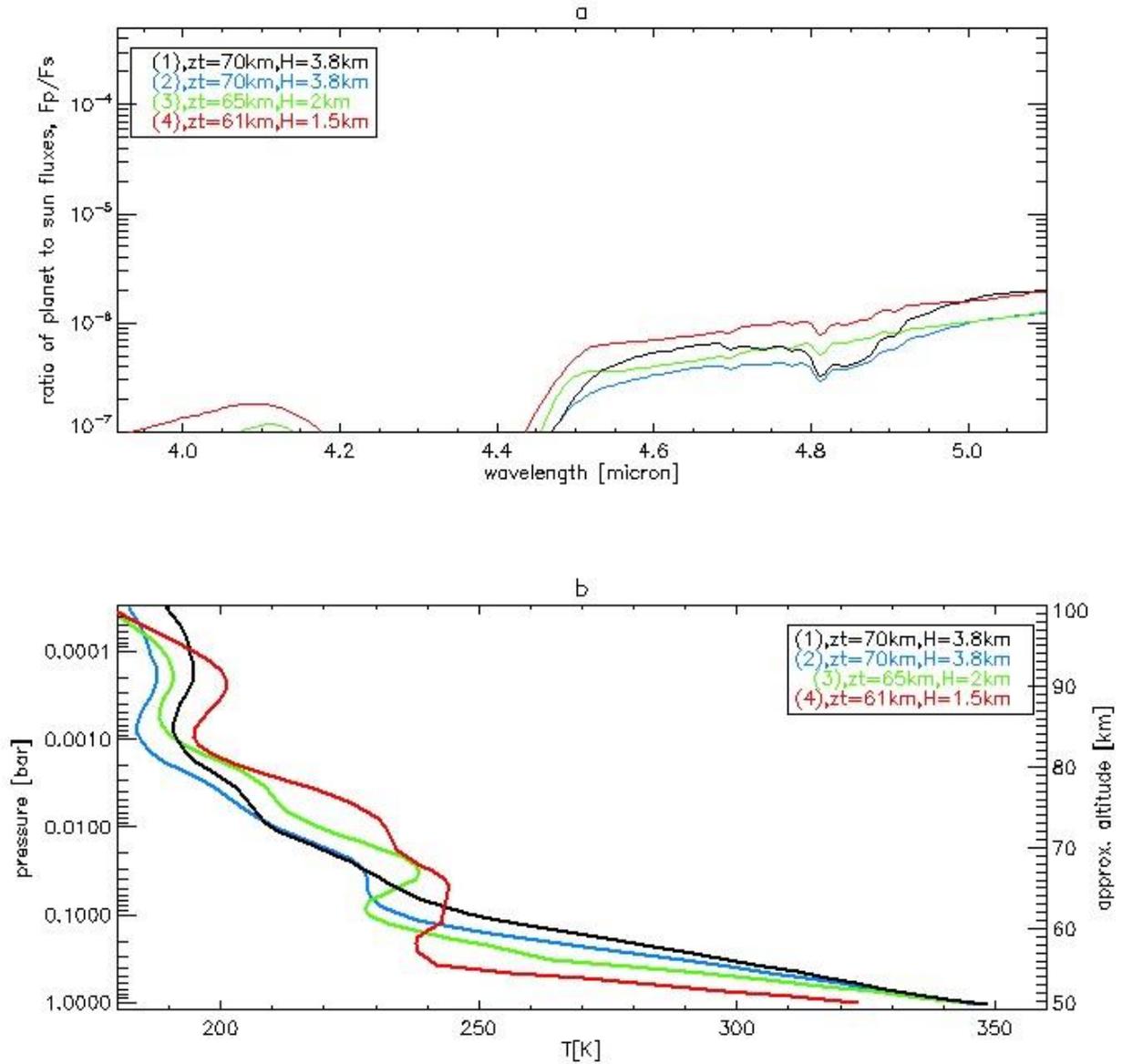

Fig.1. (a) emission spectra calculated for (b) 4 different temperature profiles and assumption on clouds: 1) $zt = 70$ km and $H = 3.8$ km (black and blue line), 2) $zt = 65$ km, $H = 2$ km (green line), 3) $zt = 61$ km, $H = 1.5$ km (red line).

The radiation from the spectral range from 3.9 to 5.1 micron will come from the mesosphere of our template super-Venus because in this spectral range clouds composed of $H_2SO_4$ will absorb the radiation coming from below the cloud level. We distinguish two absorption bands by $CO_2$ at 4.3 and 4.8 μm. We assume that a single temperature profile represents the temperature for the whole hemisphere of planet. The adopted temperature profiles are realistic Venus temperature profiles from Wolkenberg et al. (2018).

With the expected sensitivities of future facilities like JWST and Ariel, we would be able to identify only the signal beyond 5 μm (Fig.1). The largest spectral coverage is found for the red curve in Fig.1a and the temperature profile also plotted in red in Fig.1b. From the assumption on cloud properties we can derive that their top altitude is 61 km, which is below the thermal inversion seen in Fig.1b (red line). As a result, we can detect top altitudes of clouds and thermal inversions studying the signal at 5 μm and beyond.

We will expand this exploratory case study by investigating the effect of different assumptions ($zt$, $H$) on clouds on emission spectra for a given temperature profile. This effect will be considered with mesospheric temperatures because for example the thermal inversion between 55 – 65 km in the temperature profile (red in Fig.1b) affects the emission spectrum (red in Fig.1a) showing a flatter absorption band by $CO_2$ near 4.8 micron and a flatter spectral slope compared to the temperature profile (black, Fig.1b) without thermal inversion.

## 3. Results
### 3.1. Temperature profile with inversion

We study an effect of clouds on emission spectra for the temperature profile with inversion at 55 - 65 km (Fig.1b – red line – and 2b). Clouds are described by a scale height $H$, i.e. their compactness, and a top altitude $zt$. In our study we assume different $zt$ and $H$ values. Fig.2a presents emission spectra calculated for different $zt$ and $H$ values assuming the temperature profile shown in Fig.1b (red line) and presented in Fig.2b.

The whole $CO_2$ absorption band at 4.8 μm is almost flat for the different ranges of $H$ and $zt$ except for scale heights $H$ = 3.8 and 5 km at cloud top altitude $zt$ = 61 km. The signal increases with scale height from 3.8 km to 5 km for clouds with $zt$ = 61 km. Only the center of this absorption band is evident and distinguishable. The characteristic spectral features are also found beyond 5 μm. This is only observed for this value of $zt$. This is probably associated with the thermal inversion at 55 – 65 km, as only clouds with less compactness allow to sound the thermal structure deeper.

The emission spectra are then very similar for $zt$ = 61 km and 65 km. There is no impact of different scale heights $H$ on spectra for the top altitude $zt$ = 65 km. A decrease of signal in the band 4.8 μm is observed for $zt$ equal to 70 km with respect to the smaller $zt$ values. No changes are found in emission spectra in the center of the main absorption band by $CO_2$ at 4.3 μm at these cloud top altitudes ($zt$ = 61, 65 and 70 km). Different scale heights $H$ for zt = 70 km have a negligible effect on emission spectra. A more evident decrease of signal is observed for $zt$ = 75 km with different scale heights. In this case only the center of the band at 4.8 μm is pronounced along with the main absorption band by $CO_2$ at 4.3 μm. The main absorption band at 4.3 μm becomes flatter for higher cloud top altitudes than for lower values of $zt$. The significant differences between spectra are found for different scale heights $H$ and $zt$ = 85 km. Here we also observe the effect of scale heights on the spectrum. For scale heights $H$=1.5 km and 2 km, the center of the $CO_2$ absorption band at 4.8 μm completely disappears showing only the line increasing with wavelength (slope).

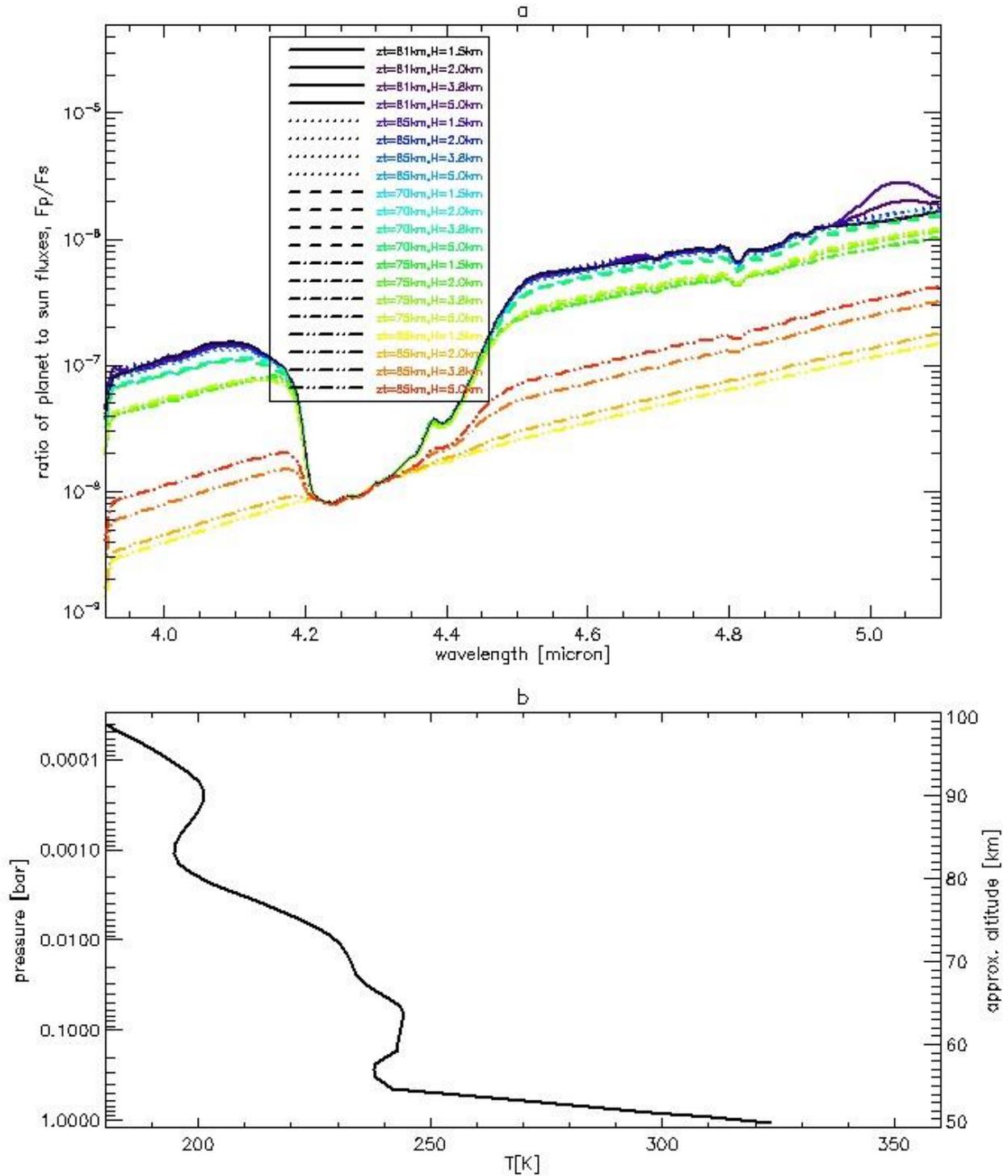

Figure 2. a) Emission spectra for temperature profile (b) (Fig.1b). Emission spectra for *zt* values from 61 to 70 km with scale heights *H* from 1.5 – 5 km are plotted by solid line. Emission spectra for *zt* = 75 km and scale heights from 1.5 – 5 km are plotted by dashed line. Emission spectra for *zt* = 85 km and scale heights from 1.5 – 5 km are plotted by dotted-dashed line.

The main absorption band at 4.3 μm is flattened so that this band is barely seen for $zt$ = 85 km and $H$ = 1.5 km. An increase of scale heights from 3.8 to 5 km leads to distinguishable absorption bands at 4.8 μm and 4.3 μm. The signal increases along with scale heights at the constant $zt$ = 85 km.

Summarizing, the effect of cloud top altitudes is more evident than that of scale heights on spectra. The density of clouds drops along with altitude. The effect of cloud scale heights is smaller than that cloud top altitudes because for example the H = 1.5 km shows indeed a high compactness of the cloud but the clouds are not completely opaque. The higher the clouds in the atmosphere, the smaller signal is observed. Less compact clouds (large scale heights) allows to detect the absorption bands despite the occurrence of high altitude clouds. For clouds less dense than H = 1.5 km (so for H = 2-5km) we will be able to characterize clouds even if the cloud top is at high altitude. The thermal inversion observed at 55 – 65 km worsens the visibility of the $CO_2$ absorption band at 4.8 μm. This absorption band is flattened due to this thermal inversion. When clouds occur below this inversion as it is for $zt$ = 61 km and scale heights 3.8 and 5 km, the absorption band is deeper than for other cloud top altitudes and scale heights.

### 3.2. Temperature profile without inversion

In this section we also consider the effect of clouds on emission spectra calculated for the temperature profile without the thermal inversion (Fig.1b – black curve, 3b). Fig.3a presents the emission spectra for different assumptions on clouds. By comparing Fig.3a with 2a significant differences can be noticed. First of all, we observe a higher signal at 4.8 μm than for the spectrum calculated for the temperature profile with the thermal inversion. The spectral features at 4.3 and 4.8 μm are more evident. The completely flat spectrum at 4.8 μm is observed for clouds with $zt$ at 85 km with different scale heights. In this case, the spectrum has the higher signal for larger scale heights (less compactness) as it was for the temperature profile with the thermal inversion. However, the signal at 5.05 μm is lower for scale heights 3.8 and 5 km and $zt$ = 85 km than the signal calculated for the temperature profile with the thermal inversion. Fig.3b presents the temperature profile at 85 km, which is similar to the isothermal line with a small inversion at the higher altitude, at 90 km. This flatness of the spectrum is not a surprise. The flattening of the absorption band at 4.8 μm begins for clouds with $zt$ at 75 km and 85 km. An increase of signal deepening the absorption band at 4.8 μm is due to increasing scale heights. This behavior is more prominent for smaller cloud top altitudes. The most pronounced absorption bands by $CO_2$ at 4.3 and 4.8 μm are observed for clouds at $zt$ = 61 km and $H$ = 5 km. The emission spectrum becomes then gradually flat with increasing cloud top altitudes. Finally, until the cloud top altitude reaches 85 km for $H$ = 1.5 km and 2 km, the emission spectrum will show the increasing line with wavelengths (slope). The signal progressively decreases with increasing cloud top altitudes.

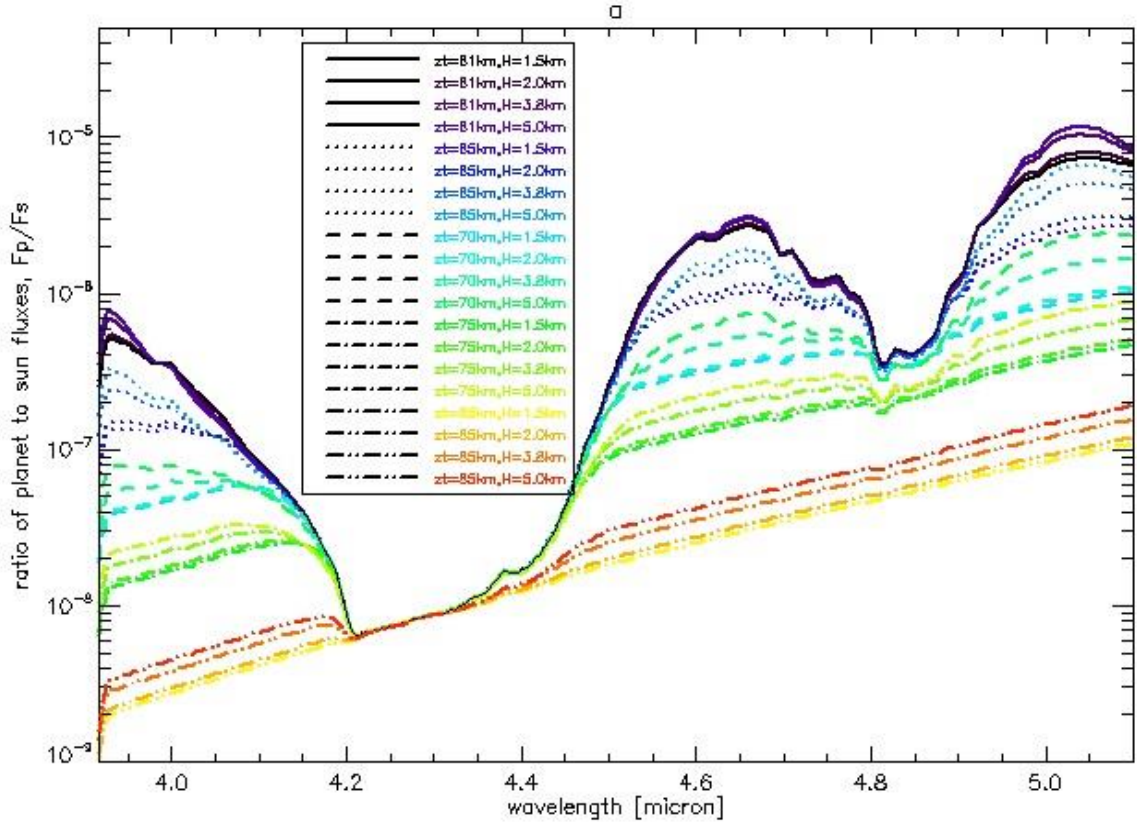

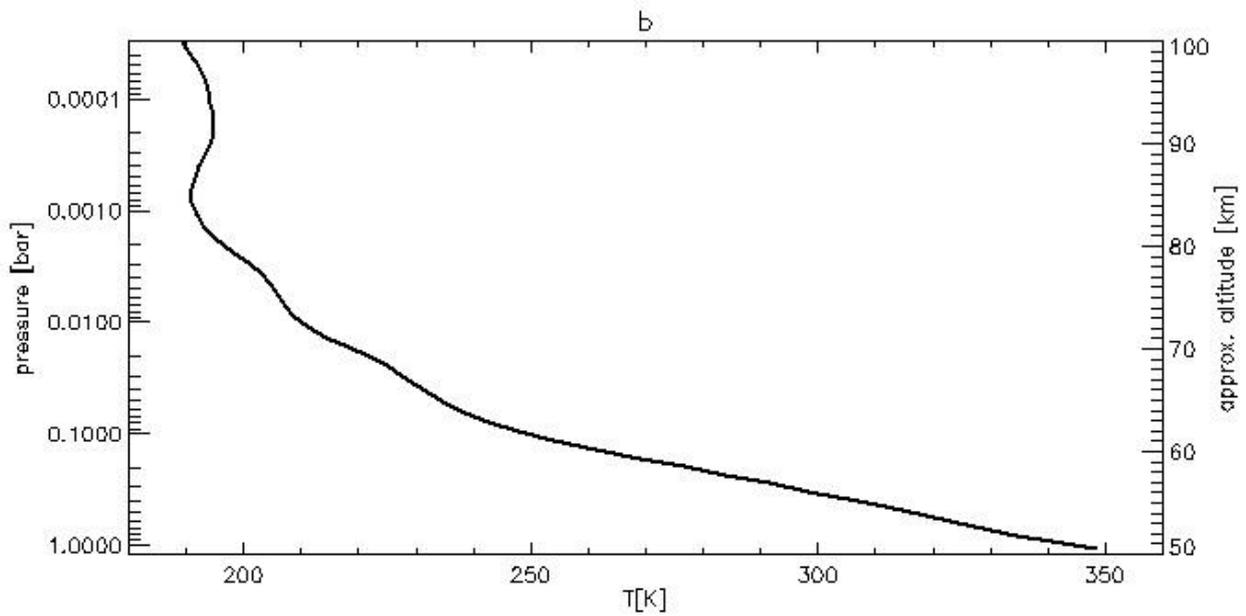

Figure 3. a) Emission spectra for temperature profile (b) (Fig.1b). Emission spectra for $zt = 61$ km and different scale heights are plotted by solid lines. Emission spectra for 65 km with scale heights from 1.5 – 5 km are plotted by dotted line. Emission spectra for $zt = 70$ km and scale heights from 1.5 – 5 km are plotted by dashed line. Emission spectra for $zt = 75$ km and scale heights from 1.5 – 5 km are plotted by dotted-dashed line. Emission spectra for $zt = 85$ km and scale heights from 1.5 – 5 km are plotted by double dotted-dashed line.

## 4. Summary and conclusions

We studied the effects of clouds on emission spectra for two different temperature profiles representing the whole hemisphere of a super-Venus planet. Our goal is to investigate the possibility to find characteristic spectral features due to clouds. We adopted two temperature profiles, one of them including a thermal inversion. Clouds are described by the scale height $H$ and the cloud top altitude $zt$. The scale height describes the compactness of clouds, thereby the larger the scale height adopted, the less dense cloud occurs.

We are able to distinguish an impact of clouds on the $CO_2$ absorption band at 4.8 μm for the two temperature profiles we considered. However, we will be unable to detect the signal for the cloud top altitude at 85 km and probably higher, and for the temperature profile without the thermal inversion in this spectral band. Surprisingly, the thermal inversion in the temperature profile helps to discern the signal coming from the cloud at 85 km of altitude. The emission signal at 4.8 μm is over 0.1 ppm for cloud scale heights equal to 3.8 and 5 km and cloud top altitude equaled to 85 km. The emission signal increases with decreasing cloud top altitudes and the $CO_2$ absorption band becomes deeper with visible characteristic spectral features. The compactness of clouds has an inverse effect on emission spectra than cloud top altitudes. Small cloud scale heights reduce the signal and the $CO_2$ absorption band becomes flat. However, the two absorption bands by $CO_2$ can give us information at which altitudes the thermal inversion and cloud occur. We know what range of altitudes (55 – 100 km) can be sounded using the absorption band at 4.8 μm and 4.3 μm. A completely flat spectrum appears for both temperature profiles with the cloud top altitude at 85 km and scale heights equal to 1.5 and 2 km. This means that clouds at high altitudes suppress or smooth the spectral features in the considered spectral range 3.9 - 5.1 μm. Then the thermal inversion is visible when the absorption band at 4.8 μm is flat but simultaneously the absorption band at 4.3 μm is distinguishable. Clouds only reinforce this effect. And finally, the effect of low altitude clouds, for altitudes lower than 85 km, can be seen in emission spectra with the two distinguishable absorption bands when the temperature profile has no thermal inversion. The cloud top altitudes and scale heights regulate the dampening of spectral features in the two absorption bands by $CO_2$.

The highest signal (10 ppm) in emission spectrum is observed between 5 - 5.05 μm for the temperature profile without the thermal inversion. The signal around 3 ppm is found at 5 - 5.05 μm for the temperature profile with the thermal inversion. Both signals are modelled with an assumption on clouds with the top altitude of 61 km and the cloud scale height equaled to 5 km. It is worth to note that when clouds occur above the thermal inversion, the $CO_2$ absorption band at 4.8 μm becomes flatter than for clouds within or below the thermal inversion. A general estimate of the sensitivity required to characterize cloudy Super Venus is about 1ppm.

By comparing signals at two spectral wavelengths, in the center of the $CO_2$ band (near 4.8 μm) and at 5 - 5.05 μm we will be able to distinguish if clouds occur in the atmosphere and at which altitudes. Small differences between signals at these two wavelengths are found for emission spectra with high altitude clouds and the thermal inversion in the temperature profile. Large differences between signals are observed for low cloud top altitudes and large cloud scale heights. These differences become smaller with an increase of cloud top altitudes. Unfortunately, the information about clouds is mixed with the thermal inversion occurring in the temperature profile. This degeneracy will be solved if we can constrain the temperature profile from a different spectral channel. The detailed analysis of emission spectra will enable to extract some information about clouds and thermal inversions in temperature profiles.

**Acknowledgements**:

P.W. and D.T. gratefully acknowledge the support of the Italian National Institute of Astrophysics (INAF) through the INAF Main Stream project ``ARIEL and the astrochemical link between circumstellar disks and planets'' (CUP C54I19000700005). D.T. also acknowledges the support of the Italian Space Agency (ASI) through the ASI-INAF contract 2018-22-HH.0.

Conflict of interest: the authors declare no conflict of interest.

Benneke et al., 2019. Water Vapor and Clouds on the Habitable-zone Sub-Neptune Exoplanet K2-18b. The Astrophysical Journal Letters, 887:L14 (9pp). https://doi.org/10.3847/2041-8213/ab59dc

Berta et al., 2013. The flat transmission spectrum of the super-Earth GJ1214b from wide field camera 3 on the Hubble Space Telescope. Astrophysical Journal, 747: 35 (17pp).

Burrows et al., 2007. Theoretical Spectral Models of the Planet HD 209458b with a Thermal Inversion and Water Emission Bands. The Astrophysical Journal, 668: L171–L174

Burrows et al., 2008. Theoretical Spectra and Light Curves of Close-in Extrasolar Giant Planets and Comparison with Data. The Astrophysical Journal, 678, 1436.

Changeat and Edwards, 2021. The Hubble WFC3 Emission Spectrum of the Extremely Hot Jupiter KELT-9b. The Astrophysical Journal Letters, 907:L22. https://doi.org/10.3847/2041-8213/abd84f

Charbonneau et al., 2002. Detection of an extrasolar planet atmosphere The Astrophysical Journal, 568:377–384.

Charnay et al., 2015. 3D modeling of GJ1214b's atmosphere: formation of inhomogeneous high clouds and observational implications. The Astrophysical Journal, 813: L1 (7pp)

Danielski et al., 2014. 0.94-2.42 micron ground based transmission spectra of the hot Jupiter HD-189733b. The Astrophysical Journal, 785: 35 (12pp)

De Kok et al., 2011. The influence of non-isotropic scattering of thermal radiation on spectra of brown dwarfs and hot exoplanets. Astronomy&Astrophysics, 531, A67

De Mooij et al., 2012. Optical to near-infrared transit observations of super-Earth GJ 1214b: water-world or mini-Neptune? Astronomy and Astrophysics, 538, A46.

De Mooij et al., 2013. Search for Rayleigh scattering in the atmosphere of GJ1214b. Astrophysical Journal, 771: 109 (7pp).

Demory et al., 2013. Inference of inhomogeneous clouds in an exoplanet atmosphere. The Astrophysical Journal Letters, 776:L25 (7pp).

de Wit et al., 2016. A combined transmission spectrum of the Earth-sized exoplanets TRAPPIST-1 b and c. Nature, vol. 537, 67. doi:10.1038/nature18641

de Wit et al., 2018. Atmospheric reconnaissance of the habitable zone Earth-sized planets orbiting TRAPPIST-1. Nature Astronomy, vol. 2, 214 - 219. https://doi.org/10.1038/s41550-017-0374-z